\documentclass[onecolumn, a4paper]{article}
\usepackage[left=3.5cm,right=3cm,top=3cm,bottom=3cm]{geometry}
\usepackage{cite}
\usepackage{amsmath,amssymb,amsfonts, }
\usepackage{algorithmic}
\usepackage{graphicx}
\usepackage{multirow}
\usepackage{url}
\usepackage{scalerel}
\usepackage{authblk}
\usepackage{color}
\usepackage[hidelinks]{hyperref}
\hypersetup{
  colorlinks   = true, 
  urlcolor     = black, 
  linkcolor    = red, 
  citecolor   = green 
}
\newcommand{\red}[1]{\textbf{\textcolor{black}{#1}}}
\newcommand{\blue}[1]{\underline{\textcolor{black}{#1}}}
\date{}

\title{\textbf{Predictive Uncertainty Estimation in Deep Learning for Lung Carcinoma Classification in Digital Pathology under Real Dataset Shifts}}

\author{Abdur R. Fayjie\thanks{Equal contribution.} $^{,1}$}
\newcommand\CoAuthorMark{\footnotemark[\arabic{footnote}]  }
\author{Jutika Borah\protect\CoAuthorMark$^{,2}$}
\author{Florencia Carbone$^{3}$}
\author{Jan Tack$^{1,3}$}
\author{Patrick Vandewalle{$^{1}$}}

\affil{$^1$KU Leuven, 3000 Leuven, Belgium
\\
$^2$Gauhati University, Guwahati 781014, India
\\
$^3$UZ Leuven, 3000 Leuven, Belgium\\
\texttt{\href{mailto:fayjie92@gmail.com}{fayjie92@gmail.com}  \hspace{0.3cm} \href{mailto:jutikaborah13@gmail.com}{jutikaborah13@gmail.com}  \href{mailto:patrick.vandewalle@kuleuven.be}{patrick.vandewalle@kuleuven.be}
}
}
\begin{document}

\maketitle
\begin{abstract}
Deep learning has shown tremendous progress in a wide range of digital pathology and medical image classification tasks. Its integration into safe clinical decision-making support requires robust and reliable models. However, real-world data comes with diversities that often lie outside the intended source distribution. Moreover, when test samples are dramatically different, clinical decision-making is greatly affected. Quantifying predictive uncertainty in models is crucial for well-calibrated predictions and determining when (or not) to trust a model. Unfortunately, many works have overlooked the importance of predictive uncertainty estimation. 
This paper evaluates whether predictive uncertainty estimation adds robustness to deep learning-based diagnostic decision-making systems. We investigate the effect of various carcinoma distribution shift scenarios on predictive performance and calibration. We first systematically investigate three popular methods for improving predictive uncertainty: Monte Carlo dropout, deep ensemble, and few-shot learning on lung adenocarcinoma classification as a primary disease in whole slide images. Secondly, we compare the effectiveness of the methods in terms of performance and calibration under clinically relevant distribution shifts such as in-distribution shifts comprising primary disease sub-types and other characterization analysis data; out-of-distribution shifts comprising well-differentiated cases, different organ origin, and imaging modality shifts. While studies on uncertainty estimation exist, to our best knowledge, no rigorous large-scale benchmark compares predictive uncertainty estimation including these dataset shifts for lung carcinoma classification.
\end{abstract}

\begin{figure}[ht]
 \centering
 \begin{tabular}{ c @{\hspace{2pt}} c @{\hspace{2pt}} c @{\hspace{2pt}} c @{\hspace{2pt}} c}
     \includegraphics[width=.19\columnwidth]{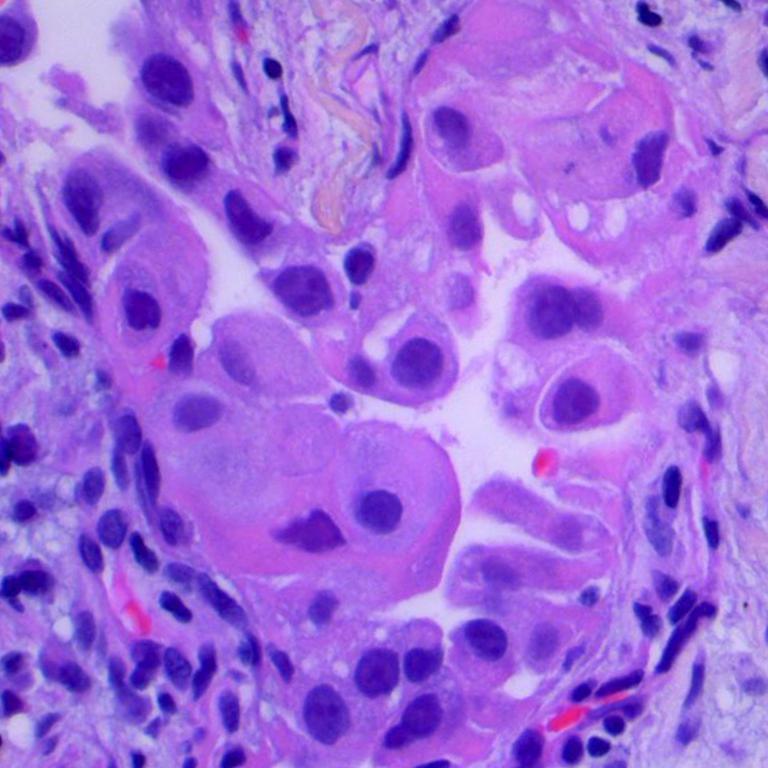} &
     \includegraphics[width=.19\columnwidth]{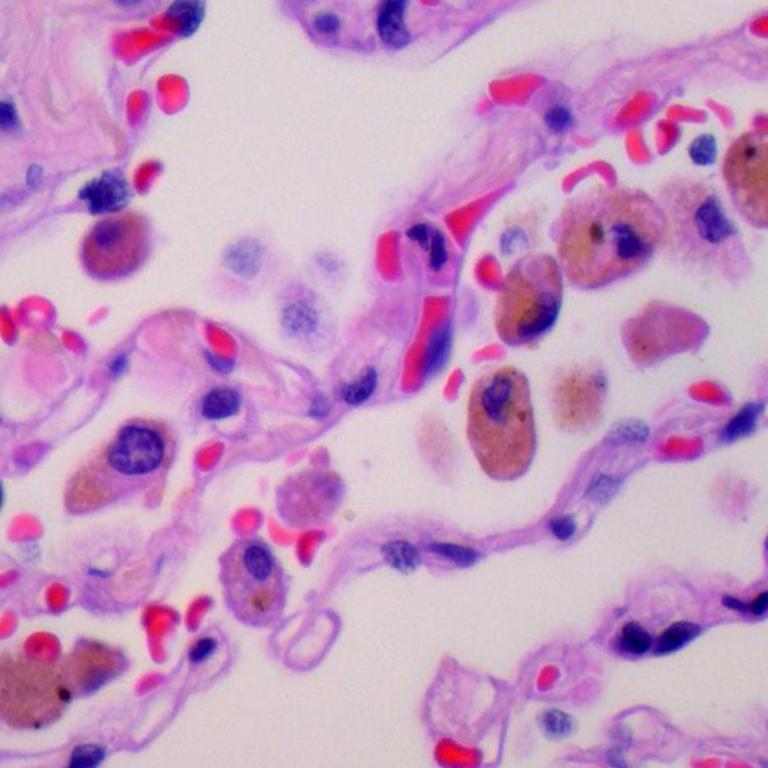} &
     \includegraphics[width=.19\columnwidth]{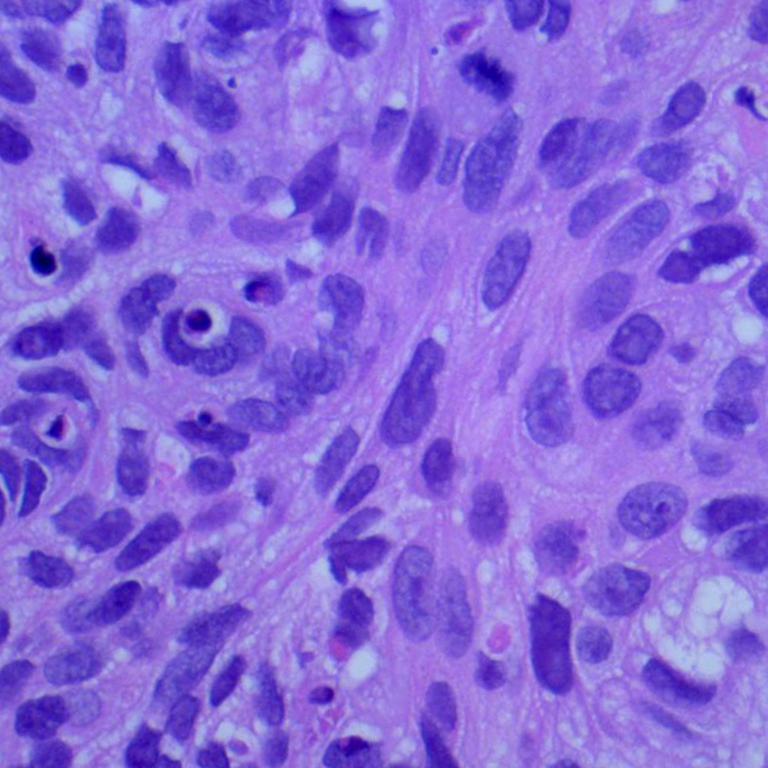} &
     \includegraphics[width=.19\columnwidth]{lungn2.jpeg} &
     \includegraphics[width=.19\columnwidth]{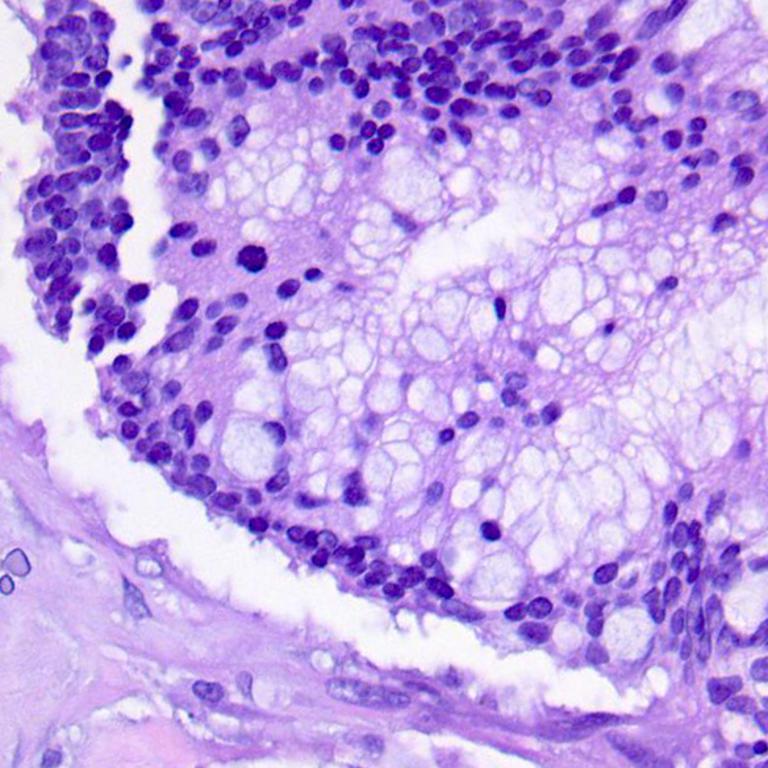} \\
   \small (a)  &
   \small (b)  &
   \small (c)  &
   \small (d)  & 
   \small (e)  \\
     \includegraphics[width=.19\columnwidth]{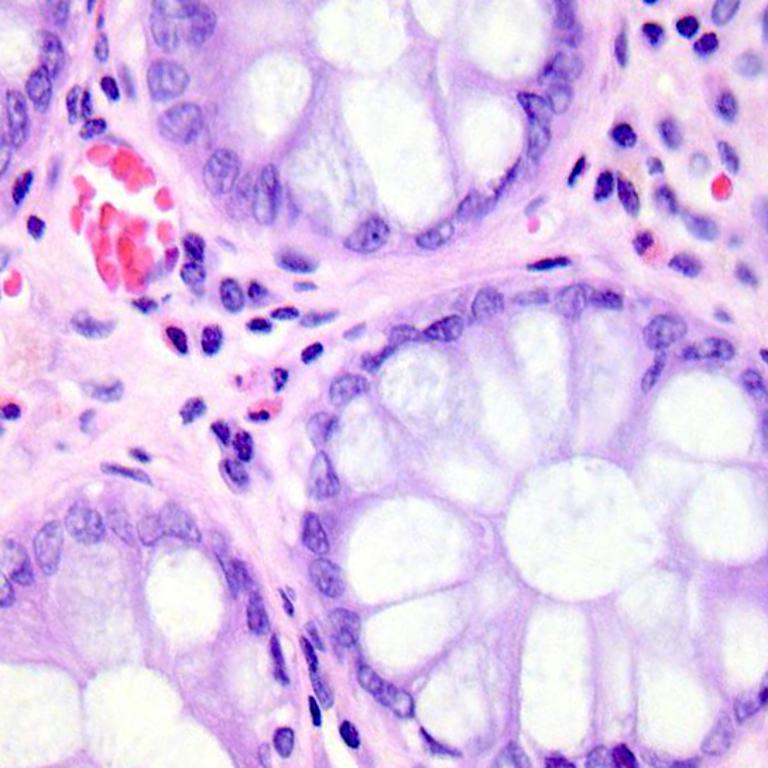} &
     \vstretch{1.132}{\includegraphics[width=.19\columnwidth]{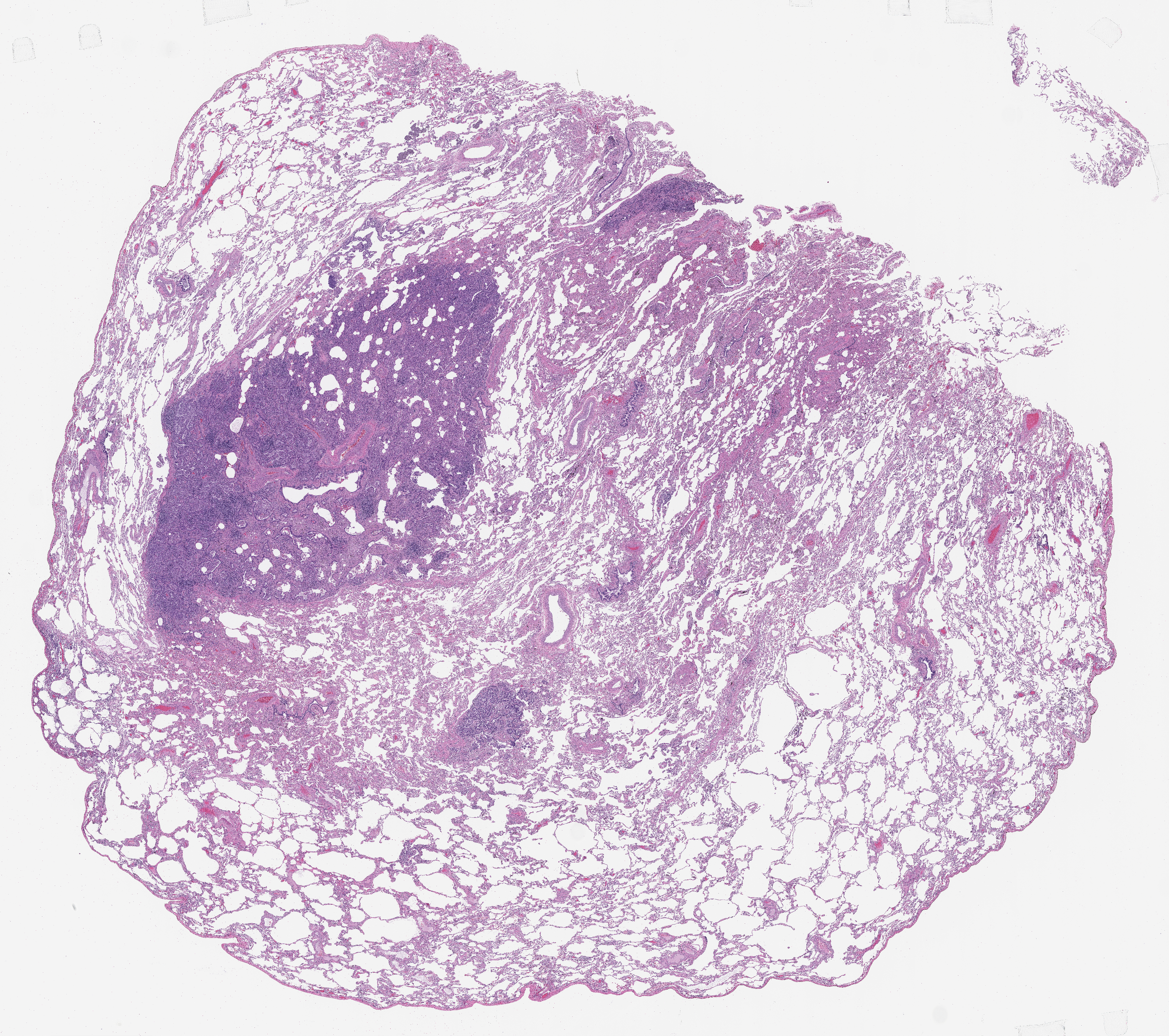}} &
     \vstretch{1.525}{\includegraphics[width=.19\columnwidth]{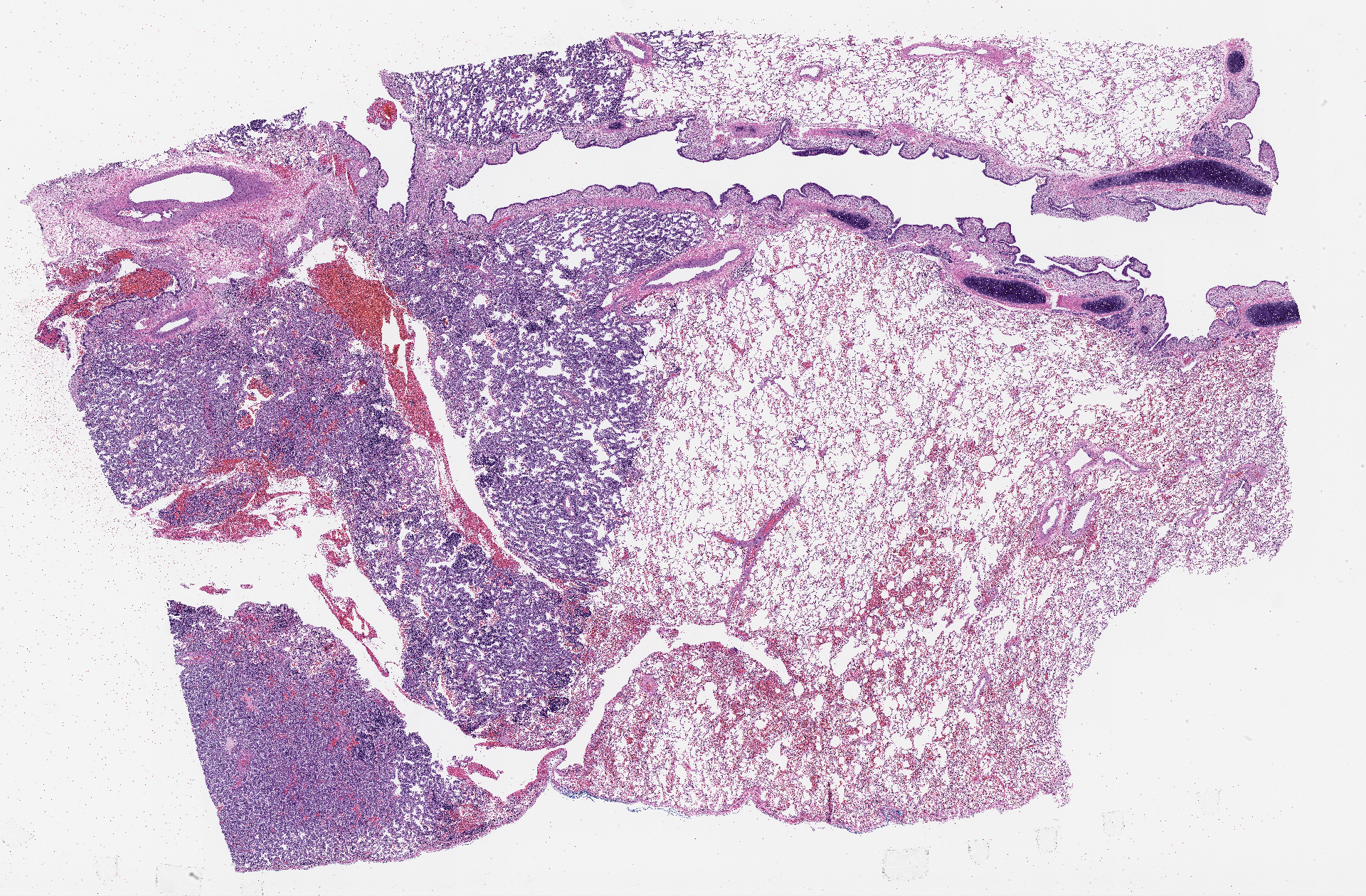}}&
     \vstretch{1.22}{\includegraphics[width=.19\columnwidth]{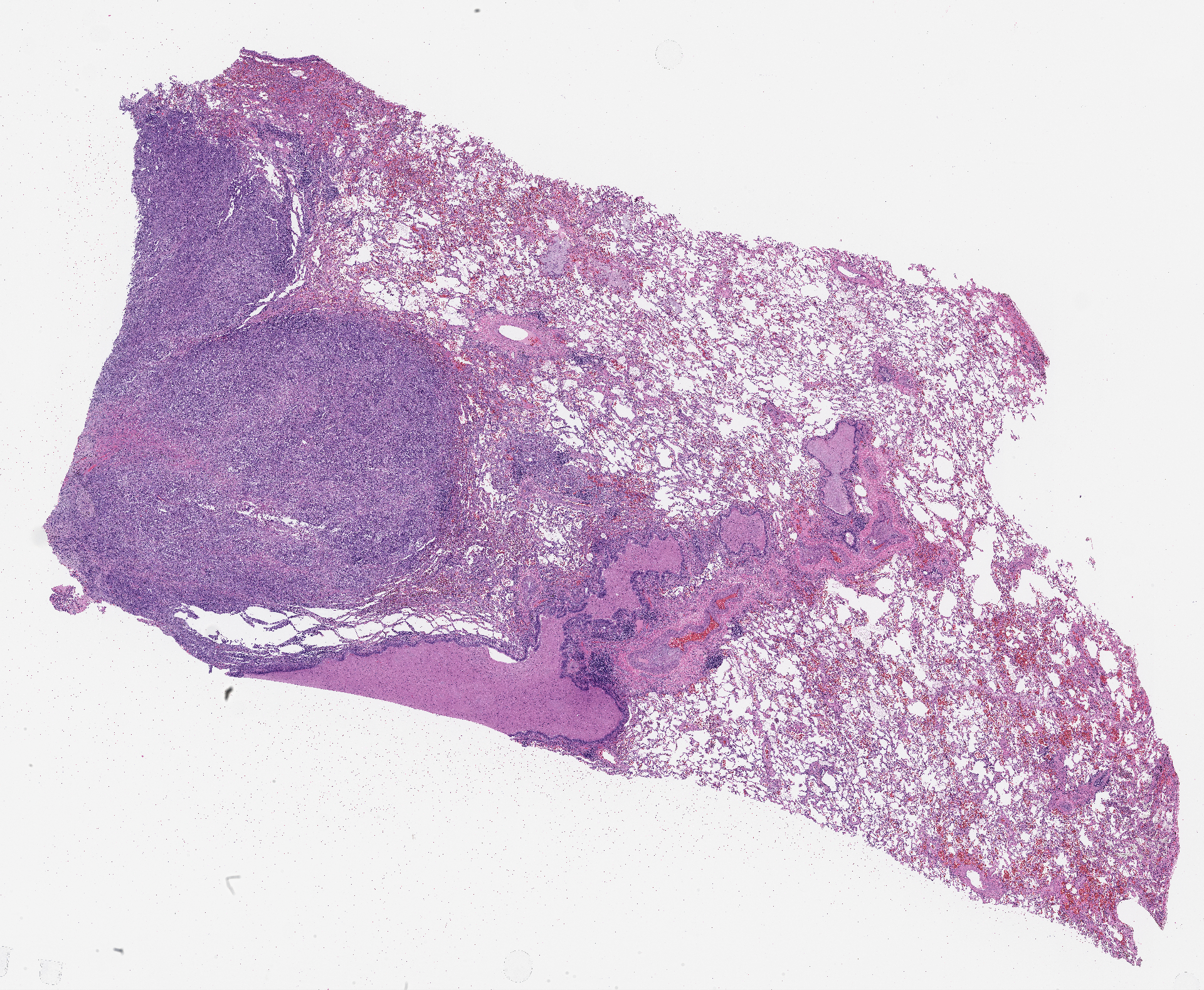}} &
     \vstretch{0.965}{\includegraphics[width=.19\columnwidth]{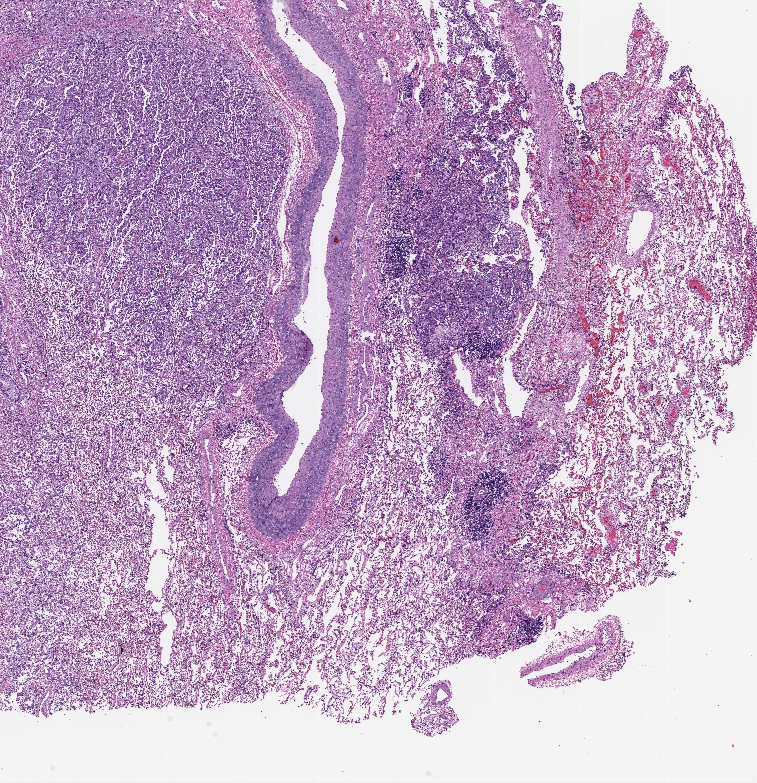}} \\
   \small (f)  &
   \small (g)  &
   \small (h)  &
   \small (i)  &
   \small (j)  \\
     \vstretch{1.28}{\includegraphics[width=.19\columnwidth]{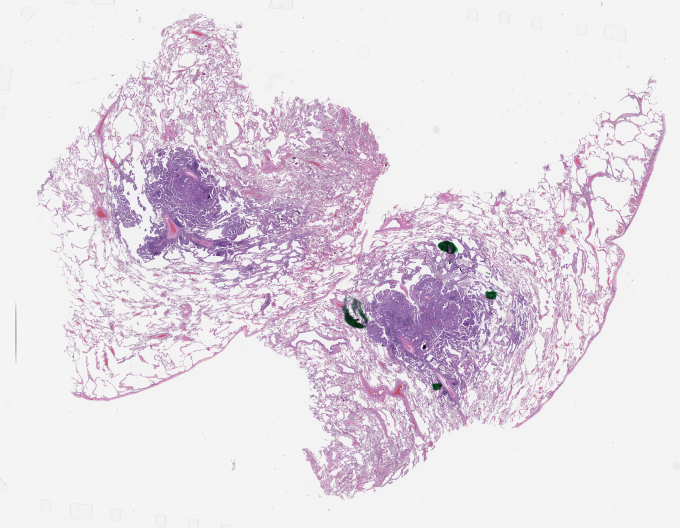}} &
     \vstretch{1.19}{\includegraphics[width=.19\columnwidth]{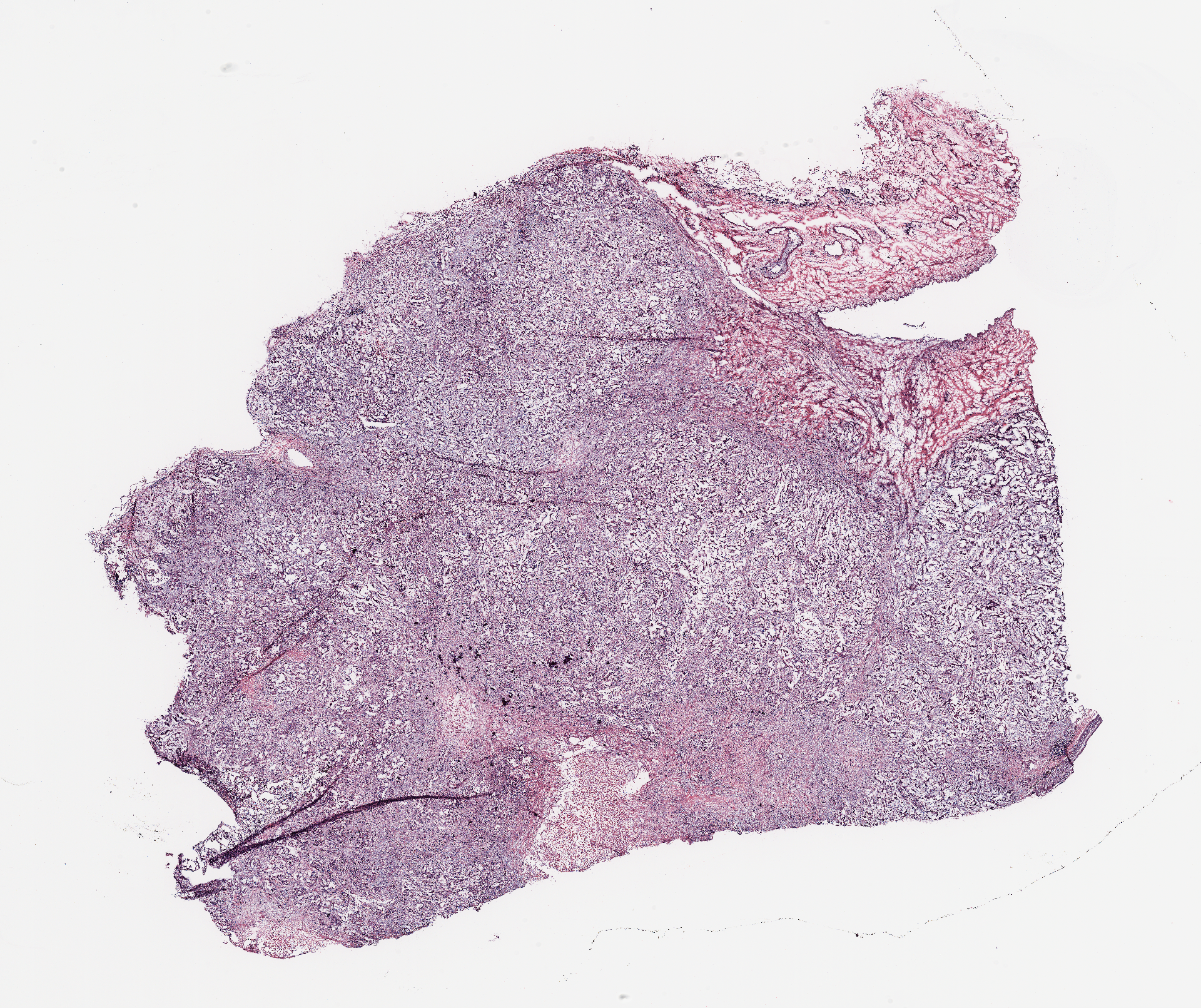}} &
     \vstretch{1.09}{\includegraphics[width=.19\columnwidth]{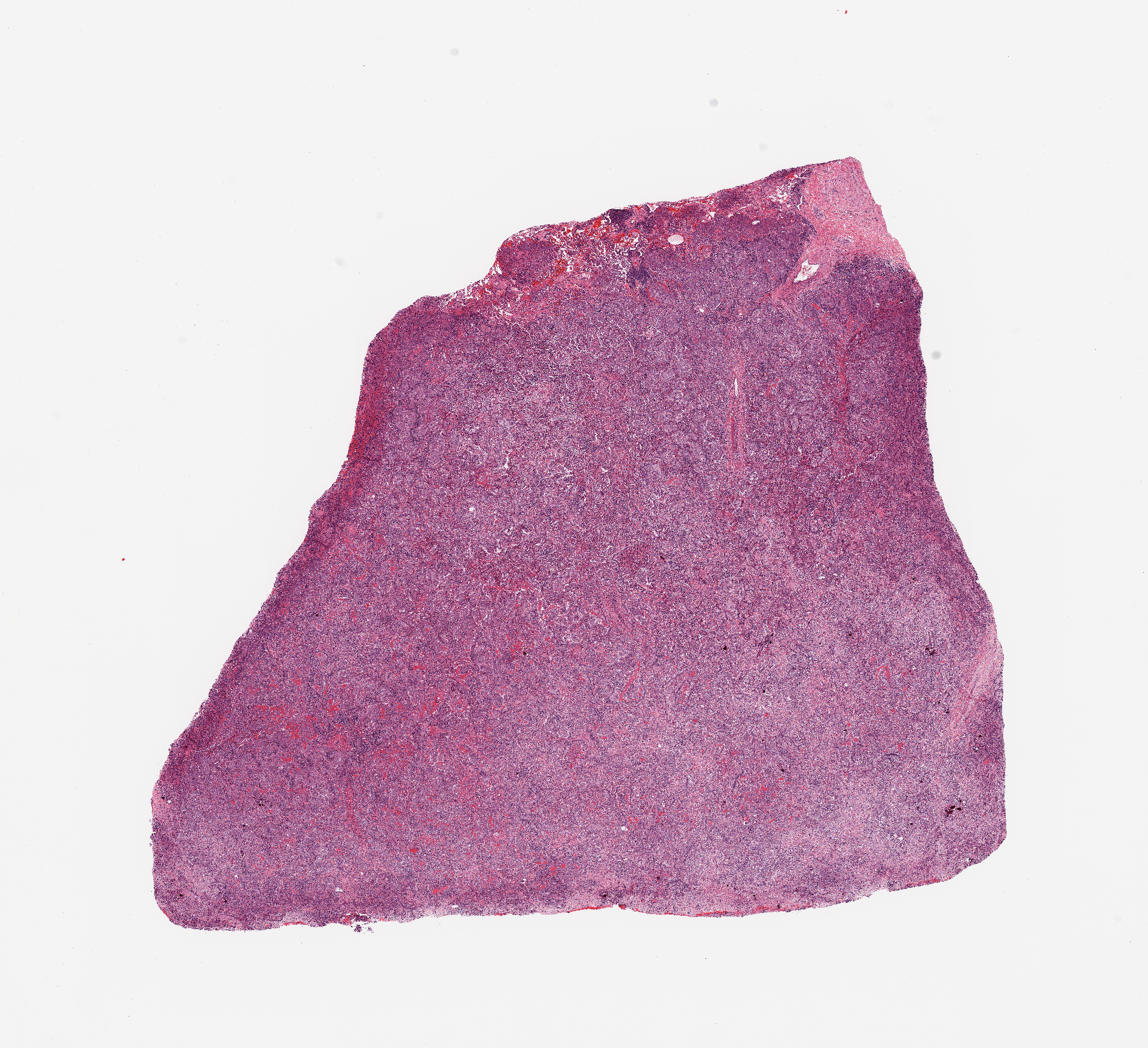}} &
     \vstretch{1.14}{\includegraphics[width=.19\columnwidth]{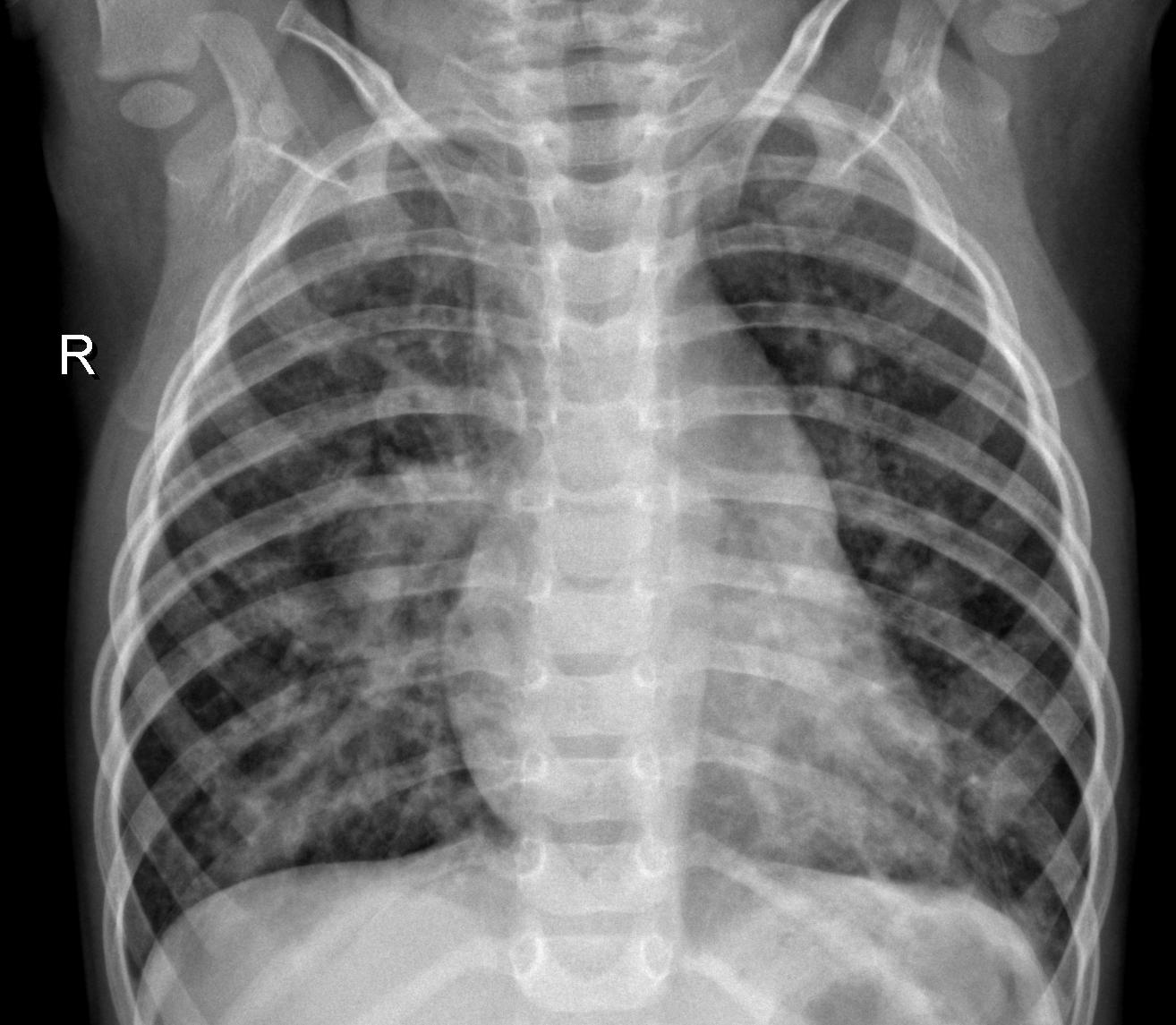}} &
     \vstretch{1.11}{\includegraphics[width=.19\columnwidth]{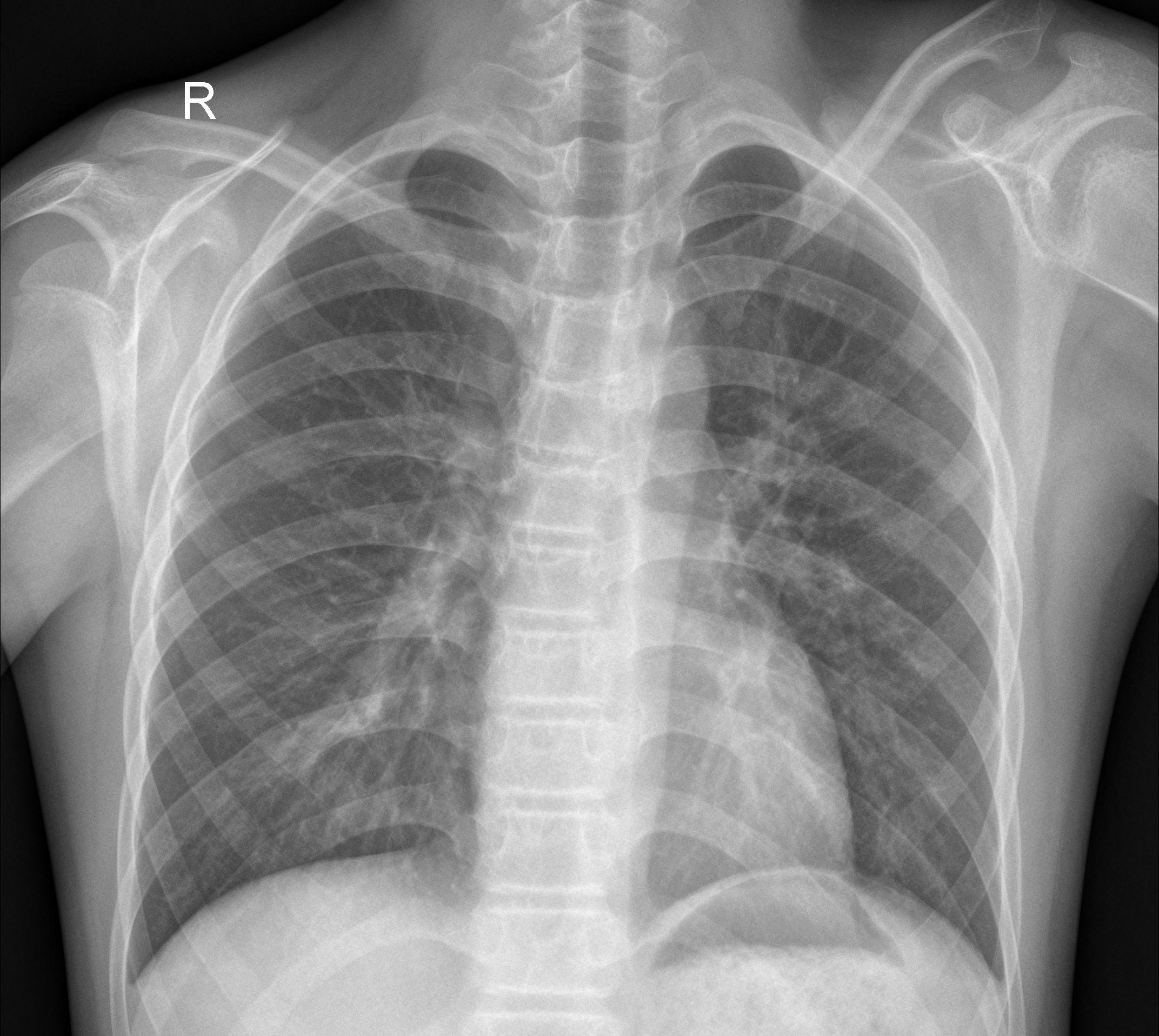}} \\
   \small (k)  &
   \small (l)  &
   \small (m)  &
   \small (n)  &
   \small (o)  \\
 \end{tabular}
  \caption{Example images from each dataset contributing to different data distribution. From the top left: WSI LC25000 (a) Lung Adenocarcinoma, (b) Normal, (c) SCC, (d) Normal SCC  (e) Colon Adenocarcinoma (f) Colon Normal; WSI BMRIDS: (g) Acinar, (h) Lepidic, (i) Solid (j) micropapillary (k) papillary; CPTAC-LUAD: (l) LUAD-positive (m) LUAD-Negative; Pneumonia CXRs: (n) Pneumonia-positive (o) Normal}
  \label{fig1}
\end{figure}

\section{Introduction} 
\label{sec:introduction}
Recent advancement across a variety of domains (security, autonomous driving, healthcare, etc.) has led to extensive utilization of deep neural networks (DNNs) in real-world applications. Modern deep learning (DL) has achieved great success in predictive accuracy for supervised learning tasks but fails to provide information on the reliability of predictions. Moreover, most recent works neglect the need to quantify uncertainty prevalent in the networks while evaluating the networks.

The performance of models drops when encountered with distributional shifts from training data, one of the reasons being overconfidence in predictions. Distributional shifts refer to the problem where the model is evaluated on test data drawn from a different distribution than the training distribution \cite{wb1, unpub1}. State-of-the-art methods fail to fill this gap between robustness and shifts in data distribution \cite{unpub2}. Many recent works focus on increasing robustness algorithmically by incorporating synthetic corruption and perturbations \cite{alKhalil2023, singh2024}. However, at test time the performance of the models drops significantly. This poses a limitation as the test samples are just modifications of the original samples at the pixel level. An ideal classifier should be robust against real-world distribution shifts as new unforeseen events may appear far from synthetically generated data over time. Ideally, with a completely different distribution this reduction in accuracy becomes more sound, where predictive distribution coincides with high entropy. An increased difference between modalities such as chest X-ray and whole slide microscopy images raises the difficulty in model generalization.

A model being overconfident in incorrect predictions can be harmful in sensitive and safety-critical applications such as medicine and healthcare \cite{A1}. Proper quantification of predictive uncertainty during test time is crucial in clinical decision-making systems to avoid significant treatment errors. The models need to demonstrate robustness against unseen shift scenarios that represent diverse changes in the distribution of clinical data over time \cite{A2}.  Distribution shifts in healthcare may exist along the axes of institutional differences (e.g.\, staffing, instruments, and data collection workflows), epidemiological changes (e.g.\, diseases, catastrophic events), temporal shifts (e.g.\, policy changes, changes in the clinician or patient behaviors over time), and differences in patient demographics (e.g.\, race, sex, age, socioeconomic background, and types of presenting illness and comorbidities). While sophisticated algorithms offer advantages in decision-making, a critical assessment of the underlying model's uncertainty is essential to guarantee the reliability and quality of predictions. Significant diversity between training and testing distributions is sufficient to deteriorate the performance of the model, negatively affecting decision-making \cite{A9}.

Towards generalization on out-of-distribution (OOD) data, the quality of predictive uncertainty raises a concern.  Models' sensitivity to such changes in data can be visualized by generating varied predictions as a base for estimating uncertainty. \textit{\textbf{Bayesian Neural Networks}} (BNNs) \cite{neal2012bayesian} can give good uncertainty estimation but are computationally expensive. Based on Bayesian approximation, a variety of methods, such as \textit{\textbf{Monte-Carlo dropout}} (MC-dropout) \cite{A10}, and  \textit{\textbf{dropout}}-based variational inference \cite{A15} have been developed for estimating predictive uncertainty in DNNs. Non-Bayesian approaches, such as \textit{\textbf{ensembling}} \cite{A12} and temperature-scaling \cite{A14} are also proposed. Moreover, in the recent decade, \textit{\textbf{few-shot learning (FSL)}} \cite{snell2017prototypical, sung2018learning, vinyals2016matching, ravi2016optimization} has emerged which aims to build robust DL models with sample efficiency, capable of generalization to new classes with a handful of labeled samples in test-time. In the context of image classification, FSL models classify a test image (query sample) belonging to a novel category, previously unseen during training by leveraging a small set of labeled images (support samples) from that category. Drawing inspiration from human cognitive abilities to adapt to new tasks or environments with limited supervision using previously acquired knowledge, FSL addresses continuously changing environments characterized by noises. The inherent data scarcity in FSL settings, especially when faced with distributional shifts, necessitates quantifying uncertainty associated with these models. 

\textbf{\textit{Contribution}}. This paper contributes to the understanding of DL and predictive uncertainty estimation for digital pathology lung carcinoma classification.  We provide a thorough and large-scale evaluation of predictive uncertainty estimation methods in neural networks. First, our experiments explore lung adenocarcinoma in tissue whole slide images (WSI) with hematoxylin and eosin staining as primary disease and focus on three main distribution shift scenarios: (i) internal test distribution, (ii) in-distribution shift (sub-types of adenocarcinomas such as lepidic, acinar, solid, papillary, and micropapillary; proteogenomic characterization analysis data: proteomic data),  (iii) OOD shift representing completely different distribution (sub-type of lung carcinoma: squamous cell carcinoma, organ shift: colon adenocarcinoma, and shift in imaging modality: Pneumonia chest X-ray mimicking clinically realistic distribution outside the training distribution. Secondly, we investigate the effect of three methods (i) \textit{\textbf{MC-dropout}}, (ii) \textit{\textbf{deep ensemble}}, and (iii) \textit{\textbf{FSL}} using entropy as uncertainty metric to evaluate generalization capability of models over clinically relevant distribution shifts. 

\textbf{\textit{Novelty.}}  The impact of distributional shifts on the performance of deep learning models in digital pathology and medical imaging remains an under-explored area. While uncertainty estimation has been previously studied, to the best of our knowledge, this work is the first to investigate its usefulness for lung carcinoma classification in digital pathology. We present a rigorous large-scale benchmark that compares various predictive uncertainty estimation methods under controlled distributional shifts. This work emphasizes the importance of understanding model uncertainty, risk, and trust, particularly as real-world data often deviates from the training distribution. Distributional shifts in this context arise from challenges in capturing the full spectrum of carcinoma types from histopathological analysis to more specific proteomic analysis, including adenocarcinoma and its sub-types. Uncertainty estimation holds particular promise in robust model design for rare conditions where training data is limited.

\section{Background} 
\subsection{Related Work}
Estimation of uncertainty is an important topic in DL research that holds the potential to provide enhanced calibrated predictions and increased robustness of neural networks. Bayesian Neural Networks are dominant in the estimation of predictive uncertainty as they often give a good uncertainty estimation by computing the parameters over the posterior distribution, given a training distribution \cite{A23, A25, A26}. However, exact inference of Bayesian networks is hard and computationally expensive, which raises the need for solutions that can deliver quality uncertainty estimates with minor modifications to the standard training pipeline. Estimation of uncertainty with modifications to softmax confidence scores \cite{unpub5}, and slight modification to neural network architectures \cite{unpub8} are new lines of research that may produce reasonable estimates of uncertainty. 

In medical imaging, most DL applications have utilized MC-dropout and deep ensemble as primary methods of estimating uncertainty. Deep ensemble \cite{A12} is arguably the simplest and most widely used method, where multiple networks are trained individually and their predictions are averaged during inference. The uncertainty methods were evaluated by discarding a certain portion of the most uncertain predictions and comparing the effect on false positive and true positive rates. Ensembling has demonstrated enhanced predictive accuracy using an ensemble of networks and has shown to give good distribution estimates of predictive uncertainty thereby improving models' performance \cite{A24}. 

Interpretation of the ensemble seems more plausible, particularly in the scenario where the dropout rates are not tuned based on the training data. Linmans et al.\ evaluate the performance of uncertainty estimation with multi-head CNNs and deep ensemble for the detection of OOD breast cancer lymphoma in sentinel lymph nodes \cite{A21}. Nair et al.\ demonstrated that the MC-dropout method can improve performance in multiple sclerosis detection and segmentation \cite{A22}. 

Different from the above methods, temperature scaling, a post-processing method, learns a scaling parameter on a validation set, but its performance is limited under distributional shifts \cite{A16, unpub4}.
Even though optimization for the temperature parameter is computationally efficient, this method fails to facilitate feature learning. Mixup \cite{A15} combines random pairs of images and their labels during training. It has recently been shown to improve the calibration of DNNs \cite{A29}.

In digital pathology, for breast cancer metastases detection in lymph nodes, Thagaard et al.\ evaluated the performance of uncertainty estimation on MC-dropout, ensemble, and mixup methods in combination and domain shifts \cite{bs1}. 
Pocevičiūtė et al.\ evaluated uncertainty estimation effectiveness for breast cancer detection in the histology of lymph nodes considering three methods: MC-dropout, deep ensembles, and test-time augmentation \cite{A20}. They considered the challenges of detecting cancer sub-types. The reported results demonstrate that deep ensembles obtained the best performance followed by test-time augmentation.

Despite significant advancement in FSL, uncertainty quantification for model evaluation remains a largely under-explored area, as evidenced by limited research works in text \cite{A28, he2023clur} and image \cite{A27, zhu2023uncertainty, wang2022uncertainty} classification. Particularly, in medical imaging, where labeled data is scarce, several FSL models have been developed for classification. However, these models overlook the aspect of uncertainty quantification. Notably, ADNet++ \cite{hansen2023adnet++} presents the sole FSL model to leverage uncertainty maps for predictions in a one-step, multi-class medical volume segmentation task. While conventional FSL focuses on generalization across different class distributions within the same source domain, performance tends to deteriorate when encountering OOD data during testing. In this context, uncertainty quantification in medical image analysis with FSL remains largely unexplored,  presenting a compelling avenue for future research.

\begin{figure}[t]
\centerline{\includegraphics[trim=2cm 6cm 1.5cm 3cm, width=1\columnwidth]{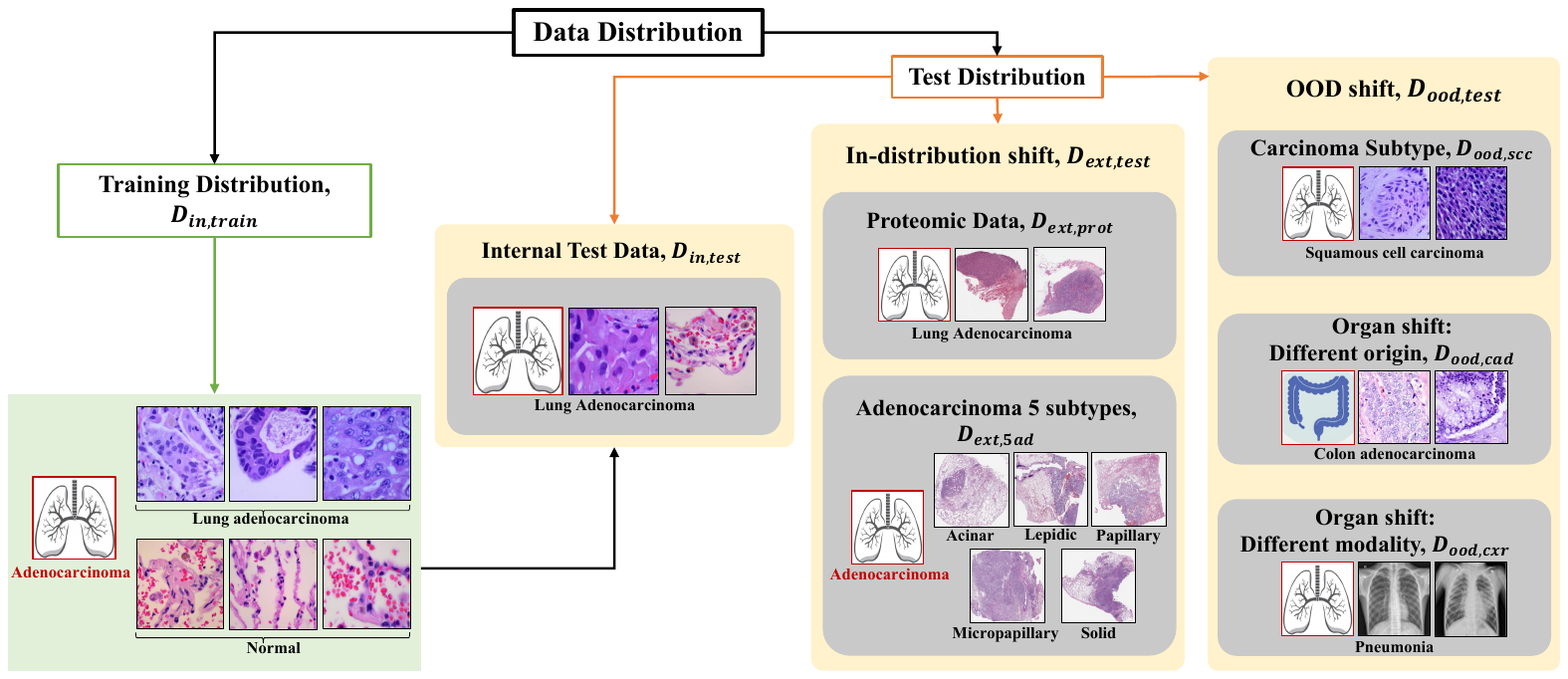}}
\caption{Experimental set-up features various distribution shifts from histopathology analysis to more specific characterization such as proteomic analysis for lung carcinoma and its sub-types classification. Training distributions, $D_{in, train}$ contains samples from LC25000 datasets. Internal test distribution, $D_{in, test}$, are taken from the same training distribution. Unseen test distribution comprises of in-distribution $D_{test, ext}$, and OOD shifts $D_{ood, test}$. The in-distribution shifts comprise two datasets with different geographical origins and characterization $D_{ext, prot}$, and class distribution $D_{ext, 5ad}$. OOD shifts consist of datasets with different carcinoma sub-type $D_{ood, scc}$ which are morphologically different, different organ origins $D_{ood, cad}$, and test imaging modality completely different from training sample modality $D_{ood, cxr}$.}
\label{fig2}
\end{figure}

\subsection{Problem Set-up}
In our image classification task, we denote an image as $x \in H\times W\times 3$ and its corresponding targets (ground truth label) as $y \in \{1, ..., C\}$, a discrete distribution over $C$ given classes. $H$ and $W$ represent the height and the width of the image, respectively. We define an internal dataset, $D_{in} = \{(x_i, y_i)\}_{i=1}^N$ that contains $N$ independent and identically distributed (\textbf{\emph{i.i.d.}}) samples. Furthermore, $D_{in}$ is sub-divided into a training subset, $D_{in, train} = \{(x_i, y_i)\}_{i=1}^{N_{train}}$ and a test subset, $D_{in, test} = \{(x_i, y_i)\}_{i=1}^{N_{test}}$ containing $N_{train}$  and $N_{test}$ \textbf{\emph{i.i.d.}} samples from $D_{in}$, respectively. 
We denote a deep neural network as $f_\theta(\cdot)$ with learnable parameters $\theta$ which model the probabilistic predictive distribution $p(y | x)$ over the targets. The classification task aims to train the network on $D_{in, train}$ to be able to make accurate predictions for a test image, $x$ as $\hat{y} = f_\theta(x)$, where $x$ is drawn from \textbf{\emph{i.i.d.}} $D_{in, test}$.
We assume that our test samples follow the same distribution as the training distribution but are independent to obtain the posterior probabilities. Note that, ${D}_{in,test}$ contains the same $k$ classes as $D_{in,train}$ seen during training.

To study a pertinent application in digital pathology and medical imaging, we consider the primary target task as the classification of lung diseases. This corresponds to both histopathology (whole slide images) and radiology imaging data (chest X-rays).  For evaluating and estimating the predictive performance and uncertainty on distributional shifts at test time, we consider a rigorous and practical classification task over data distribution representing \textbf{in-distribution shifts} ($D_{test, ext}$) and \textbf{OOD shifts} ($D_{ood, test}$), where (i) $D_{test, ext}$ $\cap$ $D_{in, train}$ = $\emptyset$, and (ii) $D_{test, ext}$ $\neq$ $D_{ood, test}$. To evaluate these scenarios, we obtain diverse datasets from multiple public sources. The task of evaluating the performance is assumed by the ability of the models to provide high-quality estimates and correctly discriminate between instances without specific re-training.

\section{Materials and Methodology}

In this section, we describe three methods for uncertainty estimation 
compared in our experiments based on their practical applicability and scalability. We provide details for each method, metrics used for the classification task, and datasets used for training, and evaluation. 

\subsection{Uncertainty estimation methods}
All the methods are trained on data (images) drawn from a training distribution, $D_{in, train}$. During inference, methods are evaluated for uncertainty estimation on the $D_{ext, prot}$, $D_{ext, 5ad}$, $D_{ood, cxr}$, $D_{ood, scc}$, and $D_{ood, cad}$ datasets. Additionally, the first two methods are also evaluated on the $D_{in, test}$ dataset. However, while evaluating FSL, we do not consider $D_{in, test}$ as we draw this distribution from $D_{in, train}$, which violates the unseen data generalization condition of FSL.

\subsubsection{MC dropout} MC-dropout has been introduced as an alternative to the computationally expensive Bayesian probabilistic method for estimating uncertainty. A neural network, $f_{\theta}(\cdot)$, parameterized by $\theta$ is trained on a training (data) set. Dropouts are generally enabled during training to regularize the learning and avoid overfitting of the neural networks. Most commonly, dropouts are disabled during test time to leverage all the connections of the network. However, following the work of \cite{A15}, we enable the dropouts during test time. In our setting, we compute the posterior probability distribution $p(\theta|x,y)$ over the trained network weights $\theta$ given the input sample $x$ and the corresponding ground truth $y$. To quantify uncertainty during inference, 100 forward passes are performed, and an average is taken to produce the final prediction. During training, we use dropout with a rate of 0.25.

\subsubsection{Deep ensemble} Deep ensemble is a non-Bayesian approach that involves training multiple neural networks \cite{A21, bs1} denoted as $f_{\theta_i}(\cdot)$, where $\theta$ is the weights and $i = \{1, 2, ..., n\}$ is the total number of networks in the ensemble. The trained weights, $\theta_i$ are utilized to classify an image $x$ during inference. The final prediction is obtained by aggregating the predictions from $i$ independent neural networks in the ensemble based on prediction confidence.  More specifically, we compute the probability of the input image $x$ to be in a class, $c$ as, 
\begin{equation}
p(y=c) = \sum_{i=1}^{n} p(y \mid \theta_i, x).
\end{equation}

\subsubsection{Few-Shot Learning} In few-shot classification, $C$-way $K$-shot tasks are defined within the context of a test dataset. In our setting, we use two test datasets, denoted by $D_{test, ext}$ and $D_{ood, test}$. Each task samples a query image $x$ and a support set comprised of $K$ labeled examples, denoted by $S_c = \{(x_i, y_i)\}_{i=1}^K$ for each class $c \in C$, drawn from the test dataset. Note that, $K < N_{train}.$ Given this support set, the model, $f(\cdot)$ is tasked with predicting a class label, $\hat y$ for the query image, $x$. 

We employ a metric-based FSL method using Prototypical Networks \cite{snell2017prototypical}. Given an image, $x$, a backbone network $f_\theta(\cdot)$  extracts a corresponding $m$ -dimensional feature vector, $z = f_\theta(x) \in \mathbb{R}^m$.  Class prototypes, indexed by $c$ are then calculated by averaging the embedding vectors for $K$ support samples as, $z_c = \frac{1}{|S_c|} \sum_{(x_i, y_i) \in S_c}f_\theta(x_i),$ where $|S_c|$ denotes the number of support samples belonging to the $k$-th class. The probability of classifying a query image $x$ into class $c \in C$ is then obtained by applying a softmax over distances to the prototypes:
\begin{equation}
\label{eq:class_prototype_definition}
p(y=c \mid x, \theta) = \frac{exp(d(f_\theta(x), z_c))}{\sum_{c=1}^C exp(d(f_\theta(x), z_c)),}
\end{equation}
The similarity metric, $d(f_\theta(x), z_c)$ utilizes Euclidean distance in our setting to measure similarities between query embedding and prototypes, $z_c \in C$. Furthermore, we set the number of classes, $C$, and support samples, $K$ to 2 and 5, respectively. The predictive uncertainty of a query sample, $x$ is then calculated as, $1- \max (p)$. We train our FSL model using an episodic training strategy that mimics the test time $C$-way $K$-shot task definition to sample data from the training dataset, i.e.\ $D_{(in, train)}$ during training.

\begin{table}[t]
\centering
\begin{tabular}{llll}
\hline
Dataset & Purpose & No. of samples & Sites\\
\hline
LC25000 & Training & 2322 (normal = 1448, & \multirow{2}{*}{Site A} \\
(Dataset 1) & (${D}_{in, train}$) &
lung adenocarcinoma = 874) & \\ \hline
\multirow{2}{*}{Dataset 1} & Evaluation         &  159 (normal = 99,    & \multirow{2}{*}{Site A} \\
          & (${D}_{in, test}$) & lung adenocarcinoma = 60) & \\ \hline

\multirow{2}{*}{CPTAC-LUAD} & Evaluation & 156 (normal = 81, & \multirow{2}{*}{Site B} \\
& ($D_{ext, prot}$) & lung adenocarcinoma = 75) & \\ \hline

\multirow{3}{*}{BMRIDS} & Evaluation & 46 (acinar = 8, lepidic = 7, & \multirow{3}{*}{Site C} \\
    & ($D_{ext, 5ad})$ & micropapillary = 10,  & \\
    &                  & papillary = 7, solid = 14) & \\ \hline

\multirow{2}{*}{CXR} & Evaluation &  624 (normal = 234,  & \multirow{2}{*}{Site D} \\
    & ($D_{ood, cxr}$) & pneumonia = 390) & \\ \hline

\multirow{2}{*}{Dataset 2} & Evaluation & 341 (normal = 130, & \multirow{2}{*}{Site A} \\
& ($D_{ood, scc}$) & scc = 211) & \\ \hline

\multirow{2}{*}{Dataset 3} & Evaluation & (normal = 163, & \multirow{2}{*}{Site A}\\
&($D_{ood, cad}$) & colon adenocarcinoma = 214) & \\ \hline
\end{tabular}
\caption{Data Distribution per class across different datasets contributing to the internal test set, in-distribution shifts, and OOD shifts}
\label{tab:data_distribution}

\end{table}

\subsection{Datasets}
Lung carcinoma is one of the most common causes of major cancer incidence and the second most common cause of cancer mortality worldwide \cite{petersen2011morphological}. This can be diagnosed pathologically either by a histologic or cytologic approach. Adenocarcinoma, Squamous Cell Carcinoma (SCC), and small and large cell carcinoma are four major histologic types of lung carcinoma. Adenocarcinoma is one of the most common carcinoma conditions which constitutes around 31\% of carcinoma cases. SCC is the second most common carcinoma accounting for roughly 30\% of the positive cases.  Most cancers of the breast, pancreas, lung, prostate, and colon are adenocarcinomas. Table \ref{tab:data_distribution} presents the data distribution settings used in our experiments. Moreover, we present an example from each class of the datasets in Fig. \ref{fig1}.

\textit{\textbf{LC25000}}: LC25000 is a histopathology image dataset with 25,000 color images in 5 classes \cite{unpub10}. The 5 classes are divided into separate subfolders each containing 5,000 images of histologic entities, namely, lung adenocarcinoma, lung benign tissue, lung SCC, colon adenocarcinoma, and benign colonic tissue.  The images are publicly available and are de-identified, HIPAA compliant, and validated. The images have a size of 768 x 768 pixels in jpeg file format. For training purposes, we considered only lung adenocarcinoma cases and kept the colon adenocarcinoma cases for inference. 
In our experiment, we have considered a training set comprising 2322 lung adenocarcinoma histopathology samples, a validation set of 524 samples, and a test of 159 samples. The dataset contains augmented views of samples with right and left rotations, as well as vertical and horizontal flipping. In our case, we have discarded some of the most complementary samples. We manually selected images that exhibit a certain level of dissimilarity. 

\textit{\textbf{CPTAC-LUAD}}:  CPTAC or the Clinical Proteomic Tumor Analysis Consortium released lung adenocarcinoma (LUAD) proteomic/phosphoproteomic data of patient tumors \cite{web1}. Proteogenomic analysis for the characterization of tumors has been performed for the systematic identification of proteins which is derived from cancer genome alterations and associated biological processes. The cohort of LUAD includes over 100 cases of Chinese and the remaining half with Vietnamese sub-populations. Both females and males are included in the study with an equal proportion of non-smokers and smokers. These are prospectively collected LUAD samples along with histopathologically normal tissues. All these include analysis for DNA, RNA, protein, and imaging which represents a comprehensive multi-omics dataset of lung adenocarcinoma patient samples with the protein, DNA, RNA, and imaging data. In our case, we have only considered protein imaging data with lung adenocarcinoma.

\textit{\textbf{BMIRDS}}: The dataset consists of 143 hematoxylin and eosin-stained formalin-fixed paraffin-embedded whole slide images of lung adenocarcinoma from the Department of Pathology and Laboratory Medicine at Dartmouth-Hitchcock Medical Center (DHMC) \cite{A32}. The dataset is de-identified and released with permission from the Darthmouth-Hitchcock Health Institutional Review Board. All WSI are labeled according to the consensus opinion of three pathologists for the predominant pattern of lung adenocarcinoma. There exists a heterogenous nature of lung adenocarcinoma. This dataset contains classes that have the predominant histological pattern of each WSI named: \textit{Lepidic, Acinar, Papillary, Micropapillary, and Solid}. All the images in the dataset are in .tif image format, scanned by Aperio AT2 whole slide scanner at 20X or 40X magnification and converted to Generic tiled Pyramidal TIFF format using libvips. MetaData.csv file contains the list of scanned slides, as well as their classes, magnification, and other details.

\textit{\textbf{Pneumonia Chest X-rays}}: A total of 5,856 chest X-rays (CXR) images are present in the dataset grouped into Normal and Pneumonia classes \cite{A31}. The dataset has three folders split into training, validation, and test sets. Class imbalance exists in the dataset with more pneumonia images than normal images. The training set has a total of 5,216 samples of which 3,875 images represent pneumonia cases and 1,341 images of normal cases, a validation set with a total of 16 samples with 8 samples per class contributing to pneumonia and normal cases, and a test set of 624 samples contributing to both classes. Here, we have only considered the test set for unseen case evaluation with 234 samples contributing to normal cases and 390 to pneumonia cases. The dataset was collected from retrospective cohorts of pediatric patients from Guangzhou Women and Children’s Medical Center, Guangzhou, China.

\subsection{Uncertainty metrics}
\textit{\textbf{Entropy}}: Given an input image, $x$, modeled as a discrete random variable with $C$ possible class labels, Shannon's entropy quantifies the amount of uncertainty associated with its classification:

\begin{equation}
    H(x) = - \sum_{c=1}^C p_{c} \log p_c.
\end{equation}

\subsection{Evaluation metrics} For evaluation of the results area-under-the-receiver-operating-characteristic-curve (AUROC) and area-under-precision-recall (AUPR) are considered. AUROC is the most common metric for performance evaluation in a binary classification problem. AUROC captures the trade-off between the true positive rate (TPR), also known as recall or sensitivity, and the false positive rate (FPR), also known as 1-specificity. AUPR is a performance measure for binary classification in a situation of class imbalance. High AUPR values indicate that the model is effective at identifying the positive classes without misclassifying many negatives as positives. Low AUPR indicates poor performance of the model as it struggles to maintain a good balance between precision and recall.
Additionally, we consider accuracy for examining in detail how classification performance varies with different methods. The TPR and FPR are defined as: $TPR = TP/(TP+FN)$ and $FPR = FP/(TN+FP)$. The accuracy metric is given as:

\begin{equation}
    Accuracy = \frac{TP + TN}{TP + FP + TN + FN},
\end{equation}
where $TP$, $TN$, $FP$, and $FN$ denote the true positive, true negative, false positive, and false negative scores, respectively.

\section{Experiments}
To study a relevant application of DL in digital pathology, we define the primary target task as identifying adenocarcinoma in hematoxylin and eosin (H\&E) stain tissue from lung sections. The main in-domain data in this study is lung adenocarcinoma. To enable the experiment, we obtained publicly available datasets and performed evaluations on three data distribution shifting scenarios with three uncertainty estimation methods. Fig. \ref{fig2} illustrates our experimental data distribution setting. We considered adenocarcinoma and SCC as well-differentiated cases of carcinoma. Furthermore, adenocarcinoma is sub-classified into more specific sub-types such as acinar, lepidic, micropapillary, papillary, and solid. We evaluate the methods' ability to generalize to these carcinoma types, their sub-types, as well as carcinoma with a different organ origin other than lungs. These distribution settings fall under $D_{ext, test}$ and $D_{ood, test}$, and are not a part of our training phase. Furthermore, we investigate if higher predictive uncertainties with higher entropy values are exhibited by the network for those unknown new instances. In this work, we train a vanilla neural network classifier as a baseline for the evaluation of the uncertainty method. 

\subsection{Data Distribution Shifts}

\begin{itemize}
    \item  \textbf{In-distribution shift}, $D_{ext, test}$. Here, we refer to in-distribution shifts as variations or changes that might affect the performance of the trained network. This distribution is not a part of the training distribution with subtle differences.  The network has to adapt to these shifts or changes to make accurate probability estimations. These shifts resemble clinical conditions where carcinoma originating in the same organ can have different sub-types and further analysis with other biological markers. Based on the understanding of histomorphology, the distinction between sub-types with the identification of the progress of the disease in pathology can give an indication of different abnormalities. 
    
    \begin{enumerate}
        \item  We investigate the method's ability to generalize to its sub-types where $p(y\mid x)$ is not fixed but has morphological features of adenocarcinoma that are not a part of training distribution. To enable this, we collect five sub-types of lung adenocarcinoma. Here, we considered histopathology data representative of lung adenocarcinoma and its 5 sub-types, $D_{ext, 5ad}$ (refer Fig. \ref{fig2}, $D_{5ad,ext})$. 

        \item  Shift containing $p(y\mid x)$ fixed, representing the same adenocarcinoma disease class. However, the distribution includes proteomic data analysis with completely different geographical origins (refer Fig. \ref{fig2}, $D_{prot,ext})$. This shows protein biological markers used to identify the lung adenocarcinoma condition in histopathology data. This includes the analysis of complete protein complement through separation, identification, and measurement demonstrated by a cell, genome, or tissue. Specifically, proteins between cancerous and their adjacent non-cancerous tissues get altered. These proteins contribute to a more precise classification of lung carcinoma and its specific types of alterations (refer Fig. \ref{fig2}, $D_{ext, prot})$. 
        
    \end{enumerate}

\end{itemize}

\begin{itemize}
    \item \textbf{Out-of-Distribution (OOD) shift}, $D_{ood, test}$. Here, we refer to OOD shifts as changes where ground truth labels are not one of the $c$ classes from the training distribution. In Fig. \ref{fig2}, the OOD shift is illustrated in the rightmost column. The network encounters data at test time that corresponds to a completely different distribution (previously unseen) and that is outside the scope of the training distribution. This type of distribution scenario offers challenges for the network to make accurate estimations. Primarily, under these shifts, the networks are expected to exhibit higher entropy values as the distributions deviate largely from the training distribution.  The classes over $D_{ext, test}$ and $D_{ood, test}$ are mutually exclusive. This category of shift includes:

    \begin{enumerate}
        \item Single organ (lung) and multiple conditions (carcinoma sub-type), $D_{ood, scc}$: This category corresponds to lungs as a single organ with the possibility of occurrence of different carcinoma as sub-types (SCC) referred to as multiple conditions. This shift concerns the challenge in dealing with different carcinoma sub-types, and whether uncertainty estimation could be beneficial in mimicking the scenario of identifying the possibility of the presence of other sub-types such as SCC or rare conditions. Also, adenocarcinoma (training distribution) and SCC (test distribution) are two well-differentiated cases, morphologically different where the former has glandular characteristics with mucin production while the latter exhibits keratinization and intercellular bridges with solid nested growth patterns (refer Fig. \ref{fig2}, $D_{ood, scc})$.
        
        \item Same condition (adenocarcinoma) with organ shifts (colon), $D_{ood, cad}$: Carcinoma that begins in glandular (secretory) cells are found in the tissue lining of certain internal organs. Adenocarcinoma contributes to most carcinomas of the pancreas, breast, lung, colon, and prostate. Morphological assessment of differentiation of colon and rectum carcinoma applies only to adenocarcinomas. Epithelial tumors other than adenocarcinomas are rarely encountered in the colon or rectum (refer Fig. \ref{fig2}, $D_{ood, cad})$.
        
        \item Different modalities and different lung conditions (other than cancer), $D_{ood, cxr}$: Clinicians assess CXRs for lung conditions ranging from pneumonia, edema, COVID-19 to TB. In this category, we collected samples that contain lung infections such as pneumonia, and healthy lungs that the network has never been trained and tested previously (refer Fig. \ref{fig2}, $D_{ood, cxr})$. 
    
    \end{enumerate}

\end{itemize}
\subsection{Implementation Details}

For MC-dropout, we trained a Residual Networks \cite{he2016deep}, specifically, the ResNet50 model with pre-trained ImageNet \cite{deng2009imagenet} weights and two dropout layers before the logits. Here, dropout is enabled during test time as an approximation of the uncertainty of the model. This is obtained by 50 stochastic forward passes through the neural network. The dropout layers have drop rates of 0.25 and 0.5. 

For deep ensemble, we trained 5 independent networks VGG19 \cite{simonyan2015very}, ResNet50, DenseNet121 \cite{huang2017densely}, Xception \cite{chollet2017xception}, and EfficientNetB0 \cite{tan2019efficientnet}. We employed Adam optimizer with an initial learning rate of $1 \times 10^{-3}$. Standard cross-entropy loss is used as the objective function. The learning rate is decayed by a factor of 0.1 when validation accuracy stops improving for 5 epochs. We trained each network for 300 epochs with a batch size of 32 until convergence.

For FSL, we adapt ResNet10 pre-trained on ImageNet weights. We trained the network using cross-entropy loss, defined as $\mathcal{L} = - \sum_{i=1} ^ {C \times n_q} y_i. \log(p(y_i|x_i)),$ where $p(y_i|x_i)$ is computed according to Eq. \ref{eq:class_prototype_definition}. We sampled 1000 training episodes. The model is validated on 100 episodes every 200 training iterations. The Adam optimizer is instantiated with a learning rate of $10^{-4}$, and an L2 regularization term with a weight of $5 \times 10^{-5}$ is applied. Finally, the trained model is evaluated on 20 test tasks sampled from the unseen test distribution, and report the average performance across these tasks.

\section{Experimental Results and Analysis}
In a practical clinical scenario, it is highly desirable for a decision making system to avoid being overconfident with incorrect predictions. To draw meaningful conclusions on the evaluation of uncertainty methods and metrics, we first trained a baseline model, a vanilla neural network that gives reasonable performance. We obtained interesting results. We summarize the results in Table \ref{tab:table2}, \ref{tab:table3}, \ref{tab:table4}, and \ref{tab:table5}.

\begin{table}[ht]
\centering
\begin{tabular}{lllll}
\hline
Methods & Distributions & Acc$\uparrow$    & AUROC$\uparrow$  & AUPR$\uparrow$   \\ 
\hline

\multirow{6}{*}{Baseline}   & ${D}_{in, test}$      & \red{0.9778}       & \blue{0.9813}    & \red{0.9840} \\ \cline{2-5}
                            & $D_{ext, prot}$     & 0.9230             & 0.9000           & 0.5621 \\ \cline{2-5} 
                            & $D_{ext, 5ad}$      & 0.4521             & 0.5621           & 0.4328 \\ \cline{2-5}
                            & $D_{ood, scc}$      & 0.9520             & 0.8063           & 0.7005 \\ \cline{2-5}
                            & $D_{ood, cad}$      & 0.5437             & 0.5791           & \blue{0.6068} \\ \cline{2-5}
                            & $D_{ood, cxr}$      & 0.6058             & 0.6063           & 0.3025 \\ \hline
\multirow{6}{*}{MC-Dropout} & ${D}_{in, test}$      & 0.9748             & 0.9659           & 0.9448 \\ \cline{2-5}
                            & $D_{ext, prot}$     & 0.8410             & 0.8926           & 0.8254 \\ \cline{2-5}
                            & $D_{ext, 5ad}$      & 0.5087             & \red{0.6325}     & \red{0.6189} \\ \cline{2-5}
                            & $D_{ood, scc}$      & 0.9028             & 0.9060           & 0.8912 \\ \cline{2-5}
                            & $D_{ood, cad}$      & 0.7029             & \blue{0.8613}    & 0.6925 \\ \cline{2-5}
                            & $D_{ood, cxr}$      & 0.5352             & \blue{0.4562}    & 0.3569 \\ \hline
\multirow{6}{*}{Ensemble}   & ${D}_{in, test}$      & \blue{0.9780}      & \red{0.9850}     & 0.9452 \\ \cline{2-5}
                            & $D_{ext, prot}$     & \blue{0.9330}      & \red{0.9550}     & \blue{0.9679} \\ \cline{2-5}
                            & $D_{ext, 5ad}$      & \blue{0.5330}      & 0.6023           & 0.5148 \\ \cline{2-5}
                            & $D_{ood, scc}$      & \blue{0.9610}      & \blue{0.9752}    & \blue{0.9592} \\ \cline{2-5}
                            & $D_{ood, cad}$      & \blue{0.7582}      & \red{0.8991}     & 0.7625 \\ \cline{2-5}
                            & $D_{ood, cxr}$      & \blue{0.6250}      & \red{0.7320}     & \blue{0.4658} \\ \hline
\multirow{6}{*}{FSL}        & ${D}_{in, test}$      &  \quad ---         & \quad ---        &  \quad ---      \\ \cline{2-5}
                            & $D_{ext, prot}$     & \red{0.9380}       & \blue{0.9481}    & \red{0.9281} \\ \cline{2-5}
                            & $D_{ext, 5ad}$      & \red{0.5934}       & \blue{0.6025}    & \blue{0.5933} \\ \cline{2-5}
                            & $D_{ood, scc}$      & \red{0.9613}       & \red{0.9790}     & \red{0.9799} \\ \cline{2-5}
                            & $D_{ood, cad}$      & \red{0.7587}       & 0.8996           & \red{0.7587} \\ \cline{2-5}
                            & $D_{ood, cxr}$      & \red{0.6440}       & 0.6769           & \red{0.4693} \\ \hline
\end{tabular}
\caption{Predictive Performance under dataset shift reported in Accuracy (Acc), AUROC, and AUPR metrics. Values in \textbf{bold} and \underline{underlined} indicate the best, and the second-best results obtained in each shift condition, respectively.}
\label{tab:table2}

\end{table}

\subsection{Evaluation on predictive performance under dataset shift}
Initially, we evaluate the predictive performance of the methods on the primary task of classifying adenocarcinoma from lung sections.   Table \ref{tab:table2} shows accuracy, AUROC and AUPR on six distributions including the internal test distribution. Overall the performance achieved on in-domain LC25000 adenocarcinoma data gives an AUROC of 0.9813 and AUPR of 0.9840  with baseline. This indicates that the baseline network was sufficient for this interpretation task. The drop in performance between the CPTAC-LUAD ($D_{ext, prot}$), BMRIDS ($D_{ext, 5ad}$), CXR ($D_{ood, cxr}$) and LC25000 ($D_{ood, scc}$, and $D_{ood, cad}$) dataset is consistent with other observations that a well-trained model should suffer a relative decrease in performance when there is a shift in data distribution having different geographical origin (different medical centers) with associated diversity within data in the datasets.

Investigating the performance of the models having different organ origins (i.e. colon) with similar disease morphology as in-domain data from the LC25000 dataset, we found a drop in performance on colon adenocarcinoma: 0.5791 AUROC and 0.6068 AUPR compared to 0.9663 AUROC and 0.9425 AUPR on SCC carcinoma subtype (\ref{tab:table2}). These results confirm that there is indeed a condition of data shift effect due to carcinoma types and different origins, which is in line with our assumptions. The same effect is also reflected in the accuracy score. 

The methods can achieve high predictive performance on both internal (LC25000, ${D}_{in, test}$) and in-distribution datasets (CPTAC-LUAD, $D_{ext, prot}$) except for five adenocarcinoma sub-type (BMIRDS $D_{ext, 5ad}$). All the methods performed significantly worse when there is a complete shift in distribution except for $D_{ood, scc}$. When evaluated on $D_{ood, cad}$, the baseline performed the worst while the ensemble and FSL performed better compared to MC-dropout. Interestingly, FSL has higher AUROC on $D_{ood, scc}$ compared to $D_{ood, cad}$ even though the cancer sub-type is morphologically different. On $D_{ood, cxr}$ all the methods have the worst performance compared to other distributions. In general, deep ensemble and FSL slightly outperform other methods on AUROC and AUPR.

\begin{table}[t]
\centering
\begin{tabular}{lllll}
\hline
Methods & Distributions & Entropy$\downarrow$ & AUROC$\uparrow$ & AUPR$\uparrow$ \\ 
\hline

\multirow{3}{*}{Baseline}       & ${D}_{in, test}$  & \red{0.0210}  & 0.9813  & 0.9840 \\ \cline{2-5}
                                & $D_{ext, prot}$ & 0.3225        & 0.9000  & 0.5621 \\ \cline{2-5}
                                & $D_{ext, 5ad}$  & 0.7963  & 0.5621  & 0.4328 \\ \hline
\multirow{3}{*}{MC-Dropout}     & ${D}_{in, test}$  & \blue{0.0300} & 0.9659  & 0.9448 \\ \cline{2-5}
                                & $D_{ext, prot}$ & 0.2256        & 0.8926  & 0.8254 \\ \cline{2-5} 
                                & $D_{ext, 5ad}$  & 0.5969        & 0.6325  & 0.6189 \\ \hline
\multirow{3}{*}{Ensemble}       & ${D}_{in, test}$  & 0.0312        & 0.9850  & 0.9452 \\ \cline{2-5}
                                & $D_{ext, prot}$ & \red{0.0293}  & 0.9550  & 0.9679 \\ \cline{2-5}
                                & $D_{ext, 5ad}$  & \blue{0.5914} & 0.6023  & 0.5148 \\ \hline
\multirow{3}{*}{FSL}            & ${D}_{in, test}$  & \quad ---     & \quad --- & \quad --- \\ \cline{2-5}
                                & $D_{ext, prot}$ & \blue{0.0322} & 0.9481  & 0.9281 \\ \cline{2-5}
                                & $D_{ext, 5ad}$  & \red{0.5963}        & 0.6025  & 0.5933 \\ 
\hline
\end{tabular}
\caption{Performance evaluation of predictive uncertainty under in-distribution shifts, reported in Entropy, AUROC, and AUPR. Values in \textbf{bold} and \underline{underlined} indicate the best, and the second-best results obtained in the entropy metric for each shift condition, respectively.}
\label{tab:table3}
\end{table}

\subsection{Evaluation on predictive uncertainty under distribution shifts (in-distribution and OOD shifts)}
We assume the predictions to exhibit higher uncertainty when the test data distribution shifts from the original source (training) distribution. This needs to be evaluated on data in a clinical context considering the relevant distributional shifts prevalent in real clinical applications. 

Table \ref{tab:table3} and \ref{tab:table4} summarizes the entropy, AUROC, and AUPR results under in-distribution shifts and OOD shifts with LC25000, CPTAC-LUAD, BMIRDS, and CXR data respectively, for different combinations of uncertainty methods. Here, we choose the entropy of the predictive distribution as an uncertainty metric for evaluating the quality of predictive uncertainty estimates. We assumed that the performance of a model would degrade as it predicts on increasingly shifted data, and ideally, this reduction in performance becomes sound and would coincide with increased entropy. For known classes, the entropy will be lower.

From the results in Table \ref{tab:table5}, for in-distribution shifts in $D_{ext, prot}$, the ensemble showed the least entropy while the baseline achieved the highest entropy compared to MC-dropout and FSL. In the case of $D_{ext, 5ad}$, MC-dropout again achieved the highest entropy compared to all other methods. Under OOD shift (\ref{tab:table4}), baseline showed the highest entropy value while ensemble showed the least value in case of $D_{ood, cad}$. On $D_{ood, scc}$, FSL showed the least entropy value compared to ensemble and MC-dropout. For increasingly shifted data, in our case $D_{ood, cxr}$, all the methods showed high entropy values compared to 
$D_{ood, scc}$ and $D_{ood, cad}$.

\begin{table}[t]
\centering

\begin{tabular}{lllll}
\hline
Methods & Distributions & Entropy$\downarrow$ & AUROC$\uparrow$ & AUPR$\uparrow$ \\ 
\hline
\multirow{3}{*}{Baseline}       & $D_{ood, scc}$ & 0.5642         & 0.8063 & 0.7005 \\ \cline{2-5}
                                & $D_{ood, cad}$ & 0.6858         & 0.5791 & 0.6968 \\ \cline{2-5}
                                & $D_{ood, cxr}$ & 0.8863         & 0.6063 & 0.3025 \\ \hline
\multirow{3}{*}{MC-Dropout}     & $D_{ood, scc}$ & 0.5322         & 0.9060 & 0.8912 \\ \cline{2-5}
                                & $D_{ood, cad}$ & 0.5963         & 0.8613 & 0.6925 \\ \cline{2-5}
                                & $D_{ood, cxr}$ & 0.8612         & 0.4562 & 0.3569 \\ \hline
\multirow{3}{*}{Ensemble}       & $D_{ood, scc}$ & \blue{0.3902}  & 0.9752 & 0.9592 \\ \cline{2-5}
                                & $D_{ood, cad}$ & \red{0.4772}         & 0.8991 & 0.7625 \\ \cline{2-5}
                                & $D_{ood, cxr}$ & \blue{0.5926}         & 0.7320 & 0.4658 \\ \hline
\multirow{3}{*}{FSL}            & $D_{ood, scc}$ & \red{0.3080}   & 0.9790 & 0.9799 \\ \cline{2-5}
                                & $D_{ood, cad}$ & \blue{0.4973}         & 0.8996 & 0.7587 \\ \cline{2-5}
                                & $D_{ood, cxr}$ & \red{0.5753}         & 0.6769 & 0..4693 \\ \hline
\end{tabular}
\caption{Performance evaluation of predictive uncertainty under OOD shifts, reported in Entropy, AUROC, and AUPR. Values in \textbf{bold} and \underline{underlined} indicate the best, and the second-best results obtained in the entropy metric for each shift condition, respectively.}
\label{tab:table4}
\end{table}

\subsection{Evaluation on Different Carcinoma sub-types}
Table \ref{tab:table5} summarizes the AUROC, AUPR, and FPR corresponding results for OOD detection. Here, a False Positive Rate or FPR is used in the evaluation pipeline for OOD detection as it indicates the model's ability to classify the positive instances correctly and avoid incorrectly classifying the negative instances as positive. Table \ref{tab:table5}, shows the performance comparison of methods on $D_{ood, scc}$ and $D_{ext, 5ad}$. The former represents a well-differentiated carcinoma condition with different cellular morphology and the latter represents an in-distribution shift with five sub-types of adenocarcinoma. All the methods show good predictive performance on $D_{ood, scc}$, however, most of the methods failed to recognize $D_{ext, 5ad}$ with the sub-types as unseen classes. Here, ensemble and FSL show better performance outperforming MC-dropout and baseline.

\begin{table}[t]
\centering

\begin{tabular}{lllll}
\hline
Methods & Distributions & AUROC$\uparrow$ & AUPR$\uparrow$ & FPR$\downarrow$ \\ 
\hline
\multirow{2}{*}{Baseline}       & $D_{ext, 5ad}$ & 0.5621 & 0.4328 & 0.9265 \\ \cline{2-5}
                                & $D_{ood, scc}$ (well diff.)  & 0.8063 & 0.7005 & 0.4465 \\ \hline
\multirow{2}{*}{MC-Dropout}  & $D_{ext, 5ad}$ & 0.6325 & 0.6189 & 0.8423 \\ \cline{2-5}
                                & $D_{ood, scc}$ (well diff.) & 0.9060 & 0.8912 & 0.3265 \\ \hline
\multirow{2}{*}{Ensemble}    & $D_{ext, 5ad}$ & 0.6023 & 0.5148 & 0.8699 \\ \cline{2-5}
                                & $D_{ood, scc}$ (well diff.) & 0.9752 & 0.9592 & \blue{0.1338} \\ \hline
\multirow{2}{*}{FSL}         & $D_{ext, 5ad}$ & 0.6025 & 0.5933 & 0.8659 \\ \cline{2-5}
                                & $D_{ood, scc}$ (well diff.) & 0.9790 & 0.9799 & \red{0.0186} \\ \hline
\end{tabular}
\caption{Performance evaluation and OOD detection on $D_{ood, scc}$, reported in Entropy, AUROC, and FPR. Values in \textbf{bold} and \underline{underlined} indicate the best, and the second-best results obtained in the FPR metric for each shift condition, respectively.}
\label{tab:table5}
\end{table}

\section{Discussion}
We evaluated current popular methods for predictive uncertainty on clinically relevant distribution shifts for lung carcinoma interpretation (classification) on H\&E-stained tissues from lung section in digital pathology. In line with our assumption, all methods demonstrated the ability to generalize from internal to external test data while maintaining the quality of predictive uncertainty. However, when applied to interpret adenocarcinoma from a different origin (organ) other than lungs (colon adenocarcinoma), all the investigated methods exhibited a decrease in performance compared to in-domain lung adenocarcinoma data. This behavior was also observed when evaluating the methods on different adenocarcinoma sub-types. However, compared to organ shift, the performance of all the methods on adenocarcinoma sub-type interpretation, with the same disease morphology decreased. Interestingly, the performance drop was more pronounced for interpreting adenocarcinoma sub-types with the same disease morphology compared to organ shift. Furthermore, the drop in performance in sub-type classification can also be attributed to the fact that the networks are trained on region-specific images, where LC25000 datasets comprise regions with carcinoma, whereas the DHMC sub-type dataset comprises samples with whole slides. These types of scenarios specifically need to be considered when designing deep-learning models for medical decision-making tasks.  

Experiments for OOD detection were conducted using adenocarcinoma and its sub-types, SSC as well-differentiated cases based on morphological patterns. Ensemble and FSL demonstrated superior performance addressing data shifts when SCC is considered as an unseen well-differentiated case. For further evaluation of OOD detection, we considered the case of detection of disease concerning lungs on radiological CXR images. We assumed that each specific modality provides inherent information and characteristics about a concerned problem and deep learning models should reflect these variations in their performances. In CXR, typically the structures are relatively simpler compared to WSI with well-differentiated anatomical regions such as the heart, lungs, and bones. On the other hand, WSI with staining contains detailed and complex information at a cellular level. This requires the networks to learn the intricate pattern and their subtle differences in cellular morphology, with varying tissue types. The difference in scale between the two modalities also has a significant impact on the performance. So, there is an increased difference in patterns between CXR and WSI which raises the difficulty (or ease) of feature extraction, thereby leading to complexity in model generalization with one compared to another. We evaluated whether the deep learning networks could detect these OOD changes within the data and how significantly these variations impact the classification across CXR and WSI.

Site-specific variations also play an important role in the interpretation of medical images. This may include variations in the sectioning of the tissues and their staining, variability in scanning, and resolution of images. Our experiments with internal test distribution showed strong predictive performance. On the other hand, with in-distribution and OOD shift conditions, experimental evaluations show variations in performance. In the context of generalization on dataset variations, MC dropout exhibited minimal performance degradation compared to the baseline, while the ensemble and FSL models displayed improved performance. Based on the findings, deep ensemble and FSL emerge as promising techniques in dealing with predictive uncertainty. However, the ensemble method incurs higher computational costs for both training and inference.

Uncertainty estimation in medical image interpretation provides valuable confidence scores for clinical decision-making, contributing to model interpretability. This allows clinicians and healthcare providers to identify cases with high uncertainty for further investigation and verification. Our results suggest that DL-based algorithms can potentially revolutionize digital pathology by integrating into healthcare diagnostic decision-making systems, provided they offer robust and reliable uncertainty estimates. Furthermore, our experiments demonstrate the added value of uncertainty estimation when coupled with appropriate measures and metrics for clearly interpreting the decisions made by the predictive models.

\section{Future Prospects and Challenges} 
Translation of DL to real clinical utility holds potential challenges. This includes addressing diverse needs and context of healthcare in the DL pipeline and establishing trust between clinicians and predictive model-generated outcomes. Furthermore, research volumes more often align with the necessity of academic incentives rather than patient and clinician needs. The clinicians are guided by their knowledge and intuitions to evaluate the value of patient health for translating it into actionable clinical information. They rely on identifying and understanding the features of the model that align with evidence-based medical practices. Incorporating predictive uncertainty estimation into these models can significantly enhance clinicians' understanding of the model's limitations. This, in turn, can serve as a valuable tool for the DL community in developing more trustworthy models for future clinical applications.


Different factors lead to variations in the performance of DL models across different modalities like differences in specific medical imaging tasks and data availability, image resolution and its preprocessing needs, information content, complexity in model architecture, and clinical variability. Each modality in medicine has its unique challenges. Understanding the subtle differences is crucial for building effective deep-learning solutions in medicine and healthcare including digital pathology. Considering and leveraging approaches, that incorporate such spectrum within model design, training, and evaluation can impact and enhance the performance of deep learning and its transferability in clinical applications.

To optimize patient-centered decision-making, healthcare needs a shift towards personalized risk management tailored to individual patient characteristics, medical history, and current health status. For real clinical practice, it is anticipated that rich, multi-institutional data representing patient diversity and heterogeneity in diseases and their states including patient demographics will be required. Future work should investigate the implementation of how the uncertainty of deep ensemble and FSL extensions can be introduced for enhanced calibration. The main focus would be on investigating if a combination of various uncertainty estimation methods would result in improved calibrated performance.

\section{Conclusion}

In this work, we investigated the application of predictive uncertainty estimation for deep learning models in carcinoma classification for digital pathology, accounting for realistic data distribution shifts. Our evaluations demonstrate strong performance on in-domain data distribution (training and internal test distribution). However, performance degrades as the data distributions shift towards more diversified and differentiated cases that are underrepresented. This performance degradation is particularly evident in real-world settings where encountering data from new hospitals or diverse disease sub-types is common. Based on the results obtained with reliable uncertainty estimates and under clear indication and monitoring of the clinicians, DL-based methods can be utilized in clinical practices. However, it is noted that these methods are not ready for alarming novel abnormalities.

\bibliographystyle{plain}
\bibliography{preprint}
\end{document}